\documentclass[5p]{elsarticle}

\usepackage{amssymb}

\newcommand{\degree}{^{\circ}}

\journal{Nuclear Instruments and Methods In Physics Research A}

\begin{document}

\begin{frontmatter}



\title{The observation of Gamma Ray Bursts and Terrestrial Gamma-ray Flashes with AGILE}

\author[INAF-IASF-Roma]{E.~Del
Monte\corref{cor1}}\ead{ettore.delmonte@iasf-roma.inaf.it}\cortext[cor1]{Corresponding
author}
\author[INFN-Trieste,Units]{G.~Barbiellini}
\author[INAF-IASF-Bologna]{F.~Fuschino}
\author[INAF-IASF-Milano]{A.~Giuliani}
\author[INFN-Trieste,Units]{F.~Longo}
\author[INAF-IASF-Bologna]{M.~Marisaldi}
\author[INAF-IASF-Milano]{S.~Mereghetti}
\author[INFN-Trieste,Units]{E.~Moretti}
\author[INAF-IASF-Bologna]{M.~Trifoglio}
\author[INAF-IASF-Milano]{G.~Vianello}
\author[INAF-IASF-Roma]{E.~Costa}
\author[INAF-IASF-Roma]{I.~Donnarumma}
\author[INAF-IASF-Roma,Uniroma1]{Y.~Evangelista}
\author[INAF-IASF-Roma]{M.~Feroci}
\author[ENEA-Bologna]{M.~Galli}
\author[INAF-IASF-Roma]{I.~Lapshov}
\author[INAF-IASF-Roma]{F.~Lazzarotto}
\author[Uniroma1]{P.~Lipari}
\author[INAF-IASF-Roma]{L.~Pacciani}
\author[ENEA-Roma]{M.~Rapisarda}
\author[INAF-IASF-Roma]{P.~Soffitta}
\author[INAF-IASF-Roma,Uniroma2]{M.~Tavani}
\author[INAF-IASF-Palermo]{S.~Vercellone}
\author[ASDC]{S.~Cutini}
\author[INFN-Pavia]{F.~Boffelli}
\author[INAF-IASF-Bologna]{A.~Bulgarelli}
\author[INAF-IASF-Milano]{P.~Caraveo}
\author[INFN-Pavia]{P.W.~Cattaneo}
\author[INAF-IASF-Milano]{A.~Chen}
\author[INAF-IASF-Bologna]{G.~Di~Cocco}
\author[INAF-IASF-Bologna]{F.~Gianotti}
\author[INAF-IASF-Bologna]{C.~Labanti}
\author[INFN-Roma2]{A.~Morselli}
\author[OA-Cagliari]{A.~Pellizzoni}
\author[INAF-IASF-Milano]{F.~Perotti}
\author[INAF-IASF-Roma,Uniroma2]{G.~Piano}
\author[INFN-Roma2,Uniroma2]{P.~Picozza}
\author[OA-Cagliari,Uninsubria]{M.~Pilia}
\author[Uninsubria]{M.~Prest}
\author[ENEA-Roma]{G.~Pucella}
\author[INFN-Pavia]{A.~Rappoldi}
\author[INAF-IASF-Roma,INFN-Roma2]{S.~Sabatini}
\author[INAF-IASF-Roma,INFN-Roma2]{E.~Striani}
\author[INAF-IASF-Roma]{A.~Trois}
\author[INFN-Trieste]{E.~Vallazza}
\author[INAF-IASF-Roma]{V.~Vittorini}
\author[OA-Roma,ASDC]{L.~A.~Antonelli}
\author[ASDC]{C.~Pittori}
\author[ASDC]{B.~Preger}
\author[ASDC]{P.~Santolamazza}
\author[ASDC]{F. Verrecchia}
\author[ASI,ASDC]{P.~Giommi}
\author[ASI]{L. Salotti}

\address[INAF-IASF-Roma]{INAF IASF Roma, Via Fosso del Cavaliere 100, I-00133
Roma, Italy}
\address[INFN-Trieste]{INFN Trieste, Padriciano 99,
I-34012 Trieste, Italy}
\address[Units]{Dip. di Fisica, Universit\`a di Trieste, Via Valerio 2, I-34127 Trieste, Italy}
\address[INAF-IASF-Bologna]{INAF IASF Bologna, Via Gobetti 101, I-40129 Bologna,
Italy}
\address[INAF-IASF-Milano]{INAF IASF Milano, Via E. Bassini 15, I-20133
Milano, Italy}
\address[Uniroma1]{Dip. di Fisica, Universit\`a di Roma ``La Sapienza'',
P.le A. Moro 5, I-00185 Roma, Italy}
\address[ENEA-Bologna]{ENEA C. R. ``E. Clementel'', Via don Fiammelli 2, I-40128 Bologna, Italy}
\address[ENEA-Roma]{ENEA C. R. Frascati, Via E. Fermi 45, I-00044 Frascati (Rm), Italy}
\address[Uniroma2]{Dip. di Fisica, Univ. Roma Tor Vergata,  Via
della Ricerca Scientifica 1, I-00133 Roma, Italy}
\address[INAF-IASF-Palermo]{INAF IASF Palermo, Via Ugo La Malfa 153, 90146 Palermo, Italy}
\address[ASDC]{ASI Science Data Center, Via G.\ Galilei, I-00044 Frascati
(Rm), Italy}
\address[INFN-Pavia]{INFN Pavia, Via Bassi 6, I-27100 Pavia, Italy}
\address[INFN-Roma2]{INFN Roma Tor Vergata, Via della Ricerca Scientifica, 1, I-00133 Roma, Italy}
\address[OA-Cagliari]{INAF Osservatorio Astronomico di Cagliari, loc. Poggio dei Pini, strada 54, I-09012, Capoterra (Ca), Italy}
\address[Uninsubria]{Dip. di Fisica e Matematica, Universit\`a dell'Insubria, Via Valleggio 11, I-20100 Como, Italy}
\address[OA-Roma]{INAF Osservatorio Astronomico di Roma, Via di Frascati 33, I-00040 Monte Porzio Catone (Rm), Italy}
\address[ASI]{Agenzia Spaziale Italiana, Unit\`a Segmento di Terra
e Basi Operative, Viale Liegi 26, 00198 Roma, Italy}


\begin{abstract}


Since its early phases of operation, the AGILE  mission is
successfully observing Gamma Ray Bursts (GRBs) in the hard X-ray
band with the SuperAGILE imager and in the MeV range with the
Mini-Calorimeter. Up to now, three firm GRB detections were
obtained above 25 MeV and some bursts were detected with lower
statistical confidence in the same energy band. When a GRB is
localized, either by SuperAGILE or Swift/BAT or INTEGRAL/IBIS or
Fermi/GBM or IPN, inside the field of view of the Gamma Ray Imager
of AGILE, a detection is searched for in the gamma ray band or an
upper limit is provided. A promising result of AGILE is the
detection of very short gamma ray transients, a few ms in duration
and possibly identified with Terrestrial Gamma-ray Flashes. In
this paper we show the current status of the observation of Gamma
Ray Bursts and Terrestrial Gamma-ray Flashes with AGILE.

\end{abstract}


\end{frontmatter}



\section{Introduction}

The italian small satellite mission AGILE (see
\cite{Tavani_et_al_2008} for further information) has been
designed, developed and produced having in mind the Gamma Ray
Bursts (GRBs) as one of the most important scientific objectives.
For this reason the X-ray monitor SuperAGILE (see
\cite{Feroci_et_al_2007} for a complete description of the
instrument) and the non-imaging Mini-Calorimeter (described by
\cite{Labanti_et_al_2009}) are equipped with on-board trigger
algorithms and dedicated telemetry data packets (see
\cite{Del_Monte_et_al_2007} and \cite{Fuschino_et_al_2008} for a
complete description). Similar methods are used in the ground
based analysis in order to detect GRBs.

At the time of writing, that is in two years of operations
including the Commissioning and Science Verification Phase (until
November 2007), the whole Cycle One (December 2007 -- November
2008) and more than one half of the Cycle Two (December 2008 --
November 2009), SuperAGILE localized 28 GRBs, about one per month.
SuperAGILE is based on 1-D imaging and its field of view is
composed of a central region, of $68 \degree \times 68 \degree$ in
which both coordinates are encoded with an error radius of 3', and
four wings, where only 1-D images are produced as narrow strips of
$6' \times 68 \degree$. Out of 28 GRBs, 17 have an error radius of
3' while for the remaining 11 the position is given by the 1-D
strip described above. In the same interval of time the
Mini-Calorimeter (MCAL), a non imaging and all-sky scintillator
detector, detected about one GRB per week in the energy band
between 300 keV and 100 MeV (see \cite{Marisaldi_et_al_2008} for
the description of the MCAL GRB capabilities). During the same
time span the AGILE Gamma Ray Imaging Detector (sensitive in the
energy band between 30 MeV and 30 GeV and described by
\cite{Tavani_et_al_2008}) found three firm detections (GRB
080514B, GRB 090401B and GRB 090510) above 30 MeV and two less
significant ones (GRB 080721 and GRB 081001) in the same energy
band.


\section{The observation of GRBs with AGILE}


Very little information is known about the emission of gamma rays
from GRBs. In fact only a handful of events has been detected in
the 30 MeV -- 10 GeV energy range by the Energetic Gamma Ray
Experiment Telescope (EGRET) aboard the Compton Gamma Ray
Observatory (CGRO) and the details are reported for example by
\cite{Dingus_2001} and references therein. The EGRET observations
showed the possibility of an extended emission of gamma rays,
longer than the prompt emission detected in the hard X-ray band by
the Burst And Transient Source Experiment (BATSE), including the
extremely peculiar case of GRB 940217 (\cite{Hurley_et_al_1994}),
with a 18 GeV photon detected after 1.3 hours from trigger.


The aim of efficiently observing GRBs in the gamma ray band has
been a primary importance driver in the development of AGILE. In
fact SuperAGILE and the Gamma Ray Imaging Detector (GRID) observe
the same region of the Sky, although with fields of view of
different dimensions, $\sim 1$ sr and $\sim 2.5$ sr respectively.
Consequently each GRB localized by SuperAGILE is also in the
center region of the GRID field of view, where the effective area
and thus the sensitivity are around their maximum. In order to
search for the detection of gamma rays, when a GRB is localized by
SuperAGILE, the significance of the GRID events consistent with
the burst time and position is estimated against the background.
The procedure is applied to all the GRBs localized in the GRID
field of view, for example by Swift/BAT, INTEGRAL/IBIS, Fermi/GBM
or IPN and the details of the GRID data analysis in this task will
be reported in a forthcoming paper.

\begin{figure}
  \includegraphics[width=6 cm, angle=90]{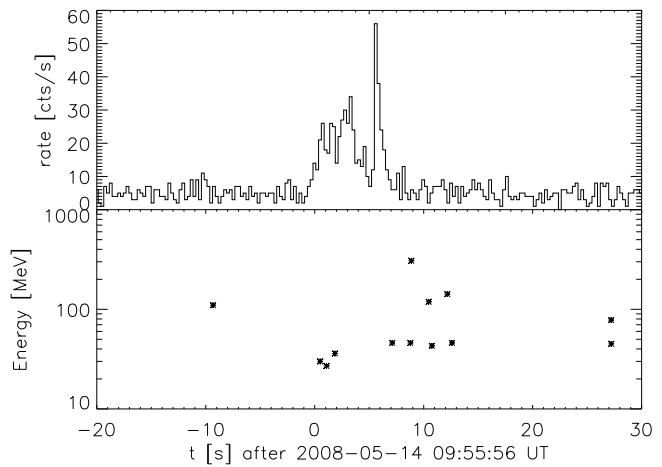}
  \caption{SuperAGILE lightcurve of GRB 080514B (upper panel) superimposed
  to the arrival time and energy of the gamma ray photons (lower panel).}
\label{fig:GRB080514B}
\end{figure}

The first GRB detected by AGILE above 30 MeV is GRB 080514B (see
\cite{Giuliani_et_al_2008} for the complete data analysis), that
is also the first burst with a significant signal in the gamma ray
band to be associated to an afterglow. The redshift of GRB 080514B
has been measured with photometric methods at $z =
1.8^{+0.4}_{-0.3}$ by \cite{Rossi_et_al_2008}. The superposition
of the SuperAGILE lightcurve and the arrival time and energy of
the photons detected by the GRID is shown in fig.
\ref{fig:GRB080514B}. From the plot it is clearly seen that the
majority of the gamma rays (about 66 \%) are emitted after the end
of the prompt emission (t0 + 7s) as detected in the hard X-ray
band, while a minority of them (33 \%) are emitted during the
prompt emission. The fluence of GRB 080514B detected by the GRID
between 20 and 50 MeV is in good agreement with the extrapolation
of the spectrum measured by Konus-Wind in the 20 keV -- 5 MeV
energy range, fitted using a Band function  (see \cite{GCN_7751}).

\begin{figure}
  \includegraphics[width=6 cm, angle=90]{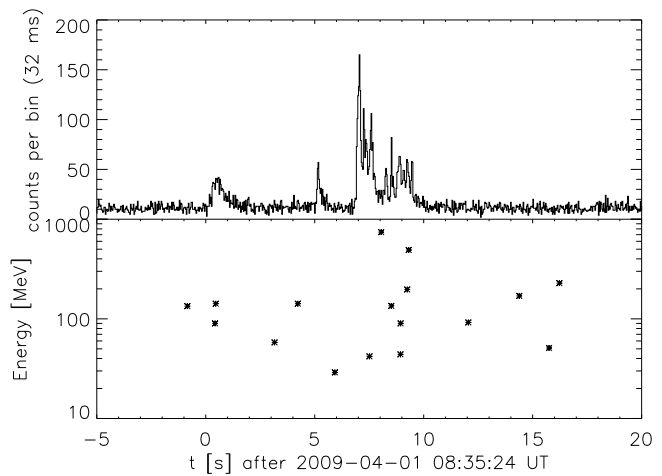}
  \caption{MCAL lightcurve of GRB 090401B (upper panel) superimposed
  to the arrival time and energy of the gamma ray photons (lower panel).}
\label{fig:GRB090401B}
\end{figure}

The prompt emission of GRB 090401B is characterized by a multipeak
structure, detected up to 2.8 MeV by MCAL (see fig.
\ref{fig:GRB090401B}) and spanning about 10 s. As shown in the
figure, in this case the majority of the photons in gamma rays (76
\%) is emitted simultaneously with the X-rays, while only a small
fraction (24 \%) is found after the end of the prompt phase.

\begin{figure}
  \includegraphics[width=6 cm, angle=90]{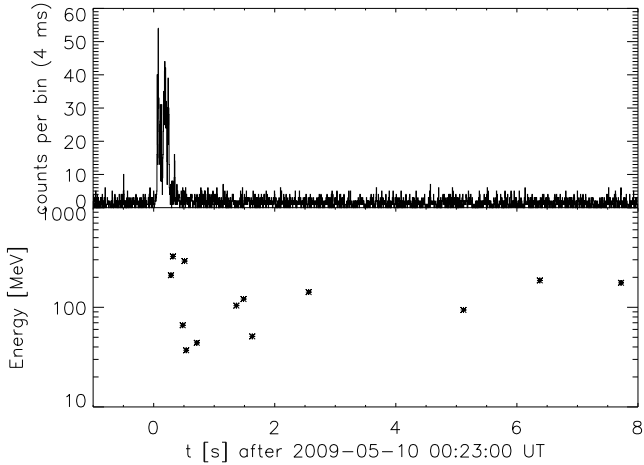}
  \caption{MCAL lightcurve of GRB 090510 (upper panel) superimposed
  to the arrival time and energy of the gamma ray photons (lower panel).}
\label{fig:GRB090510}
\end{figure}

The extended emission of gamma rays is particularly evident in the
case of GRB 090510, that is also the first short GRB with a gamma
ray emission detected by AGILE. As shown in fig.
\ref{fig:GRB090510}, the prompt emission is composed of a narrow
and hard peak, of about 200 ms duration and very clearly seen by
MCAL up to tens of MeV energy. In this case the gamma rays are not
simultaneous with the prompt emission while their detection starts
just at the end (see fig. \ref{fig:GRB090510}). GRB 090510 has
been localized by Swift/BAT (\cite{GCN_9331}) and detected also by
Fermi/LAT (\cite{GCN_9334}). The redshift has been measured
spectroscopically by VLT/FORS2 and is 0.903 (\cite{GCN_9353}).

Up to now AGILE has detected five GRBs in the gamma ray band in
around two years of observation while the Fermi satellite mission
has reported in the GCNs nine detections in about one year. If we
sum the detection rate of the two missions we obtain a total rate
of about one GRB per month in gamma rays. Since the two
instruments, GRID and LAT, have similar fields of view, we may
take into account the results of both satellites in order to
obtain an overall detection rate. It seems that the phenomenon of
the extended emission is a distinctive feature of the GRBs
emitting in the gamma-ray band. In fact the same property has been
detected, for example, by Fermi in the lightcurve of GRB 080916C
(\cite{Abdo_et_al_2009}).

\section{Terrestrial Gamma-ray Flashes}


Terrestrial Gamma-ray Flashes (TGFs) are short and intense bursts
with typical emission in the MeV region and duration of few ms,
discovered by BATSE on CGRO (see \cite{Fishman_et_al_1994}). The
spectrum is harder than that of cosmic GRBs and the incoming
direction is compatible with the Earth atmosphere. In nine years
of operations BATSE detected only 78 TGFs mainly because of
limitations in the trigger logic architecture. The sample
statistics was greatly increased by RHESSI (\cite{Smith2005}),
whose first catalogue (see \cite{Grefenstette_et_al_2009})
contains 820 TGFs detected between 2002 and 2008. The association
between TGFs and atmospheric lightning and thunderstorm activity
has been proven by \cite{Inan_et_al_2006} and
\cite{Cohen_et_al_2006} by means of temporal and spatial
correlation with lighting strokes localized by their signature at
VLF frequencies (sferics).


The observation of TGFs takes advantage of the MCAL on-board
trigger logic, that is active on several timescales ranging
between 285 $\mu$s and 8 s and allows to send to telemetry
photon-by-photon data in a time window of 60 s centered at the
trigger time (including the energy information and a time tag with
1 $\mu$s accuracy) in case a trigger is issued. It is the first
time that a timescale as short as 285 $\mu$s is used in a space
mission.



On the time scale of 16 ms or shorter, the average rate of the
MCAL detections is 6.8 triggers per orbit, corresponding to about
95 triggers per day. In order not to miss any faint event, we
decided to keep deliberately as low as possible the threshold of
the on-board algorithm and to leave the event selection to the
ground analysis, whose main purpose is the rejection of the
triggers of instrumental origin.

A substantial fraction of the instrumental triggers is related to
a particular payload status, clearly marked in the housekeeping
data, and are easily rejected. After excluding this class we apply
further criteria, based on the hardness ratio (HR) and fluence, in
order to refine the dataset and reject instrumental triggers not
identified in the previous step because of the limited timing
accuracy of the housekeeping data. The HR is computed as the ratio
between the number of events with energy higher than 1.4 MeV and
the number of them below 1.4 MeV. The cut of HR $\geq 0.5$ was
finally selected and only the triggers with at least 10 photons in
the burst time interval are considered. All the events satisfying
the criteria listed above are visually inspected to exclude
further contamination from instrumental effects and properly check
the start time and duration of the events. Following these
criteria, a total number of 34 events  are
found between June 2008 and March 2009 (corresponding to a rate of about 4
events per month), which are considered as good TGF candidates.

\begin{figure}
  \includegraphics[width=9 cm]{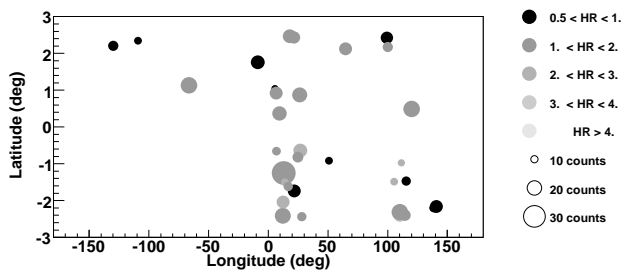}
  \caption{Geographical distribution of the AGILE TGF sample detected between June 2008 and March 2009. For each event the grey-scale indicates the HR and the marker size is proportional to the number of detected counts.}
\label{fig:TGF_geographical}
\end{figure}

The geographical coordinates of each TGF are derived from the
footprint of AGILE at the trigger time. Since AGILE is on an
equatorial orbit with $\sim 2.5 \degree$ of inclination, only a
very narrow region across the Equator is span and no high latitude
coverage can be obtained. The geographical distribution of the
MCAL dataset is shown in fig. \ref{fig:TGF_geographical}. In the
figure the clustering of the TGFs above Africa, with about half of
the events ranging in the longitude interval from $0 \degree$ to
$30 \degree$, and South-East Asia, with about one third of the
triggers in the interval between $90 \degree$ and $120 \degree$,
is clearly seen. The AGILE TGFs geographical and local time
distributions  well match those for the RHESSI TGFs, when a
consistent cut in latitude is applied.

The sum of the net exposure of all the TGFs in the MCAL sample
amounts to 51 ms. The dataset shows a total number of 47 photons
with energy higher than 10 MeV and 8 photons above 20 MeV. These
number are highly significant since, in the same energy intervals,
the expected background counts are 2.3 and 1.3 respectively. The
highest photon's energy detected is larger than 40~MeV. The
cumulative spectrum can be well fitted with a power law with
exponential cutoff, with a cutoff energy of about 9~MeV. The same
spectral model closely match also the RHESSI cumulative spectrum,
with a difference in the normalization factor of about a factor 2.
This difference in normalization may be ascribed both to
calibration issues and to dead time effects. The effects of dead
time were proven to be of particular importance for RHESSI and
BATSE as shown in \cite{Grefenstette_et_al_2008} and are still
under evaluation for what concerns MCAL. Ref.
\cite{Marisaldi_et_al_2009} reports a detailed description of the
selection criteria and the TGF candidate properties, as well as
the comparison of the AGILE sample with the RHESSI one.

%

\section{Discussion and conclusions}

The observation of GRBs by Fermi and AGILE missions is showing
that the emission of gamma rays from this class of sources is
rather uncommon, with a detection rate of about one event per
month. The continuing observation will allow to increase the
statistic of the data sample and may help to find correlations
between the detection of gamma rays and other features of the
GRBs. So far the gamma rays are detected mainly during the prompt
phase of the GRBs and usually belong to an extended emission,
lasting longer than the prompt emission in hard X-rays. The
fraction of gamma rays in the extended emission may vary,
depending on the position of the peaks in the lightcurve. Another
important property, reported in the case of GRB 080514B, is that
the fluence in gamma rays is found on the extrapolation of the
spectrum of the prompt emission detected up to the MeV region. A
similar feature has been found in the spectrum of GRB 080916C,
detected by Fermi/LAT (\cite{Abdo_et_al_2009}). The analysis of
the energetics of other gamma ray emitting bursts is still in
progress. Up to now no peculiar feature is found in the X-ray
afterglow of gamma ray emitting GRBs but the sample is still
small.



After the onset of the timescale shorter than 64 ms on the
on-board trigger of the AGILE MCAL, the instrument is detecting
a population of short and intense bursts with properties compatible with those of TGFs.
Both the geographical and local time distributions as well as the
cumulative spectrum of the AGILE TGF sample well match the same
distributions for the RHESSI sample, confirming that AGILE is
actually detecting TGFs and the goodness of the selection criteria
applied. Improvements in the selection strategy and trigger logic
are expected to increase the TGF detection rate. Thanks to the
almost equatorial orbit, AGILE can provide a continuous monitoring
of the equatorial region, especially concerning central Africa and
South East Asia, where some of the most severe TGF-producing
thunderstorms develop.




\section*{Acknowledgments}

AGILE is a mission of the Italian Space Agency, with
co-participation of INAF (Istituto Nazionale di Astrofisica) and
INFN (Istituto Nazionale di Fisica Nucleare). This work was
partially supported by ASI grants I/R/045/04, I/089/06/0,
I/011/07/0 and by the Italian Ministry of University and Research
(PRIN 2005025417). INAF personnel at ASDC are under ASI contract
I/024/05/1. The authors warmly acknowledge the support by the team
of the InterPlanetary Network (IPN).

\end{document}